%Paper: gr-qc/9312034
%From: hu@umdhep.umd.edu
%Date: Thu, 23 Dec 1993 11:33:40 EST

%%%%%%%%%%%%%%%%%%%%%%%%%%%%%%%%
\magnification=\magstep1
\baselineskip=24truept
\pageno=1

\centerline {\bf Uncertainty Relation for a Quantum Open System}
%{\bf Squeezed States and Uncertainty Relation at Finite Temperature}
\vskip .5cm
\centerline{\bf B. L. Hu $^{\ast}$}
\centerline{\it Department of Physics, University of Maryland,
College Park, MD 20742, USA}
\vskip .5cm
\centerline{\bf Yuhong Zhang ${\ast}$}
\centerline{\it Biophysics Lab, Center for Biologics Evaluation and Research, }
\centerline{\it Food and Drug Adminstration, 8800 Rockville Pike, Bethesda,
MD 20982, USA}
\vskip .5cm
\centerline{(umdpp 93-162)}
\vskip 10cm

\noindent - Submitted to Physical Review A,  Dec. 23, 1993

\noindent $^{\ast}$ Email addresses: hu@umdhep.umd.edu, zhang@helix.nih.gov
\vfill
\eject

\centerline{\bf Abstract}

We derive the uncertainty relation for a quantum open system
comprised of a Brownian particle interacting with a bath of quantum oscillators
at finite temperature.
We examine how the quantum and
thermal fluctuations of the environment contribute to the uncertainty in the
canonical variables of the system. We show that upon contact with the bath
(assumed ohmic in this paper) the system evolves from a quantum-dominated
state to a thermal-dominated state in a time which is the same as
the decoherence time in similar models in the discussion
of quantum to classical transition.
This offers some insight into the physical mechanisms involved
in the environment-induced
decoherence process. We obtain closed analytic expressions for
this generalized uncertainty relation under the conditions of high temperature
and weak damping separately. We also consider under these conditions an
arbitrarily-squeezed initial state and show how the squeeze
parameter enters in the generalized uncertainty relation.
Using these results we examine the transition of the system from
a quantum pure state to a nonequilibrium quantum statistical state and to
an equilibrium quantum statistical state. The three stages are marked by the
decoherence time and the relaxation time respectively.
With these observations we explicate the physical conditions
when the two basic postulates of quantum statistical mechanics become valid.
We also comment on the inappropriateness in the usage of the word
classicality in many decoherence studies of quantum to classical transition.

\vfill
\eject

\centerline {\bf Introduction}

This paper aims at expounding the relation of quantum and thermal
fluctuations and their relative importance in affecting the behavior of a
quantum open system [1].
The demarkation of quantum, classical and thermal regimes
is not always clearly noted, their usual definitions or usage oftentimes
are imprecise and misleading.
For example, one hears the vague identification of a high temperature
regime as the classical regime. One also hears the loose statement that
classical is defined as the regime when the Heisenberg uncertainty principle
ceases to hold. Many recent discussions of decoherence via interaction with
an environment [2] view the disappearance of the off-diagonal components of
a reduced density matrix in some special basis
or the selection or emergence of a set of classical histories [3]
as signalling a quantum to classical transition.
What are the roles played by quantum and thermal fluctuations in all these
processes and how are these issues related to each other? [4]
%A particular {\it aim} here is to rectify and make more precise
%the meaning and connotation of quantum, classical and thermal in
%their ordinary usage (See Hu and Zhang Drexel3).

To seek an answer to these questions we  focus on the derivation of a
generalized uncertainty principle valid at finite temperature.
The uncertainty in the canonical variables are caused by both the
vacuum and thermal flucatuations of the system and the bath.
We study this problem because while simple enough to yield analytic results
it also embodies these issues fully. A summary of the
main results of this investigation was recorded earlier in [5].

It is a well-known fact in quantum mechanics that a lower bound exists in
the product of the variances of pairs of noncommutative observables.
Taking the coordinate $x$ and momentum $p$ as examples, the Heisenberg
uncertainty principle states that with $(\Delta x)^2=~<x^2>-~<x>^2$,
the uncertainty function is
$$
U_0 ^{QM} =(\Delta x)^2(\Delta p)^2 \geq {\hbar^2 \over 4}
{}~~~(T=0,~~ quantum~ mechanics)
                                                              \eqno(1.1)
$$
The existence of quantum fluctuations is a verified basic physical
phenomenon. The origin of the uncertainty relation can be attributed as
a mathematical property of Fourier analysis [6] which describes quantum
mechanics as a wave theory. Recent years have seen effort in establishing
a stronger relation based on information-theoretical considerations [7].
In particular we refer to the papers of Anderson and Halliwell [8]
and Halliwell [9].

In realistic conditions quantum systems are often prepared and studied
at finite temperatures where thermal fluctuations permeate. At high
temperatures the equipartition theorem of classical statistical mechanics
imparts for each degree of freedom an uncertainty of $kT/2$.
Thus the uncertainty function for a one-dimensional harmonic oscillator
approaches the limit
$$
U_T ^{MB} \approx ({kT\over\Omega})^2
{}~~~(high~T,~~ classical~ statistical~ mechanics)
                                                                \eqno (1.2)
$$
where $\Omega$ is its natural frequency $\Omega$.
This result, obtained by assuming that the system obeys the
Maxwell-Boltzmann distribution, is  usually regarded as the classical limit.
For a system of bosons in equilibrium at temperature $T$, the application
of canonical ensemble gives the result in quantum statistical mechanics
as
$$
U_T ^{BE} = {\hbar^2 \over 4} [ \coth ({{ \hbar \Omega} \over {2 kT}})]^{2}
{}~~~(all~ T,~~quantum~ statistical~ mechanics)
                                                                \eqno (1.3)
$$
which interpolates between the two results (1.1) and (1.2) at $T=0$ and
$T>> \hbar \Omega /k$. This result applies to a system already in
equilibrium at temperature $T$.

Under equilibrium and stationary conditions the finite-temperature uncertainty
principle is just this simple. It has been studied before via thermofield
dynamics [10]. Here we
%don't want to be restricted by these conditions, but
aim to tackle the fully nonequilibrium problem.
%We also don't think it is necessary conceptually
%to depict thermal properties with a different set of basic variables.
Using the microdynamics of a quantum system as
starting point we view thermal fluctuations as statistical variations of the
coarse-grained environmental variables with which the quantum system interacts,
the exact microdynamics of the system and the environment obeying only the laws
of quantum mechanics.
%The problem under study can thus be stated equivalently
%as finding the uncertainty relation in an open quantum system.

Our  model is that
of a collection of coupled harmonic oscillators where one is distinguished as
the system of interest and the rest as bath.
This model has been studied extensively before [2][11-17].
We use the influence functional
method [12] to incorporate the statistical effect of the bath on the system.
%As the microdynamics is explicit in this approach, one can study how the
%%result
%depends on the properties of the bath and the system-bath interaction.
At time $t_0$ we put the system in contact with a heat bath at temperature
$T$ and follow its time evolution.
We consider an initial Gaussian wave packet and calculate the spread
$<\Delta x>^2, < \Delta p>^2 $ and the uncertainty function $U_T$ due to
both quantum and thermal fluctuations. The main result is given
formally by Eq. (2.3.20), which,
under the conditions of high temperature and weak damping, simplifies
to Eq. (3.3.4) and Eq. (3.4.6) respectively.
The results for an initial coherent
or minimum-uncertainty state was first reported in [5].
Anderson and Halliwell [8] have recently shown that this result is equal to
the lower bound of an information-theoretical definition
of uncertainty with different initial coherent states.  Here we derive
the uncertainty relation for the general case of an arbitrarily-squeezed
initial state where the squeeze parameter appears explicitly [17].

The paper is organized as follows: In Sec. 2, we begin with a short summary
of the influence functional formalism applied to the Brownian motion model.
(Details can be found in [12-15]. Readers familiar with this can
skip to Sec. 2.2.)
We then derive the reduced density matrix for a Gaussian wavepacket with
nonzero initial position and momentum and arrive at a finite
temperature uncertainty function (2.3.20) in terms of the elementary
functions $u_i$ and coefficients $a_{ij}$ introduced in [15].
In Sec. 3 we assume an ohmic bath and derive the forms of certain coefficients
$f_n, g_n$ related to $u_i, a_{ij}$, expressing them in terms of the
squeeze parameter $\delta$.
We then work out the uncertainty function for a closed system
(zero-coupling with the environment)
of squeezed states (3.2.5). In Sec 3.3 and 3.4 we obtain the uncertainty
function for high temperature and weak damping cases, given  respectively
by (3.3.4) and (3.4.6). For  weak couplings, if the
initial state is a minimum-uncertainty state, we recover the result (3.4.7)
obtained earlier by us [5] and by Anderson and Halliwell [8]. In Sec. 4
we use our results to discuss the relative importance of quantum and
thermal fluctuations and show how this bears on two related issues: One
on the relation of uncertainty to decoherence studies [2], and the other
on the demarkation of quantum, classical and thermal regimes.
In our
results one can identify a time where the thermal fluctuation overtakes the
quantum fluctuation. This time is identical to the decoherence time
obtained in earlier studies of quantum to classical transition [2].
The other time scale of significance is
that of relaxation time. We caution that after the decoherence time,
although the system is describable in terms of probabilities, it cannot yet
be regarded as classical because of the spin-statistics effect,
and it remains in a state of nonequilibrium.  After the relaxation time
the two basic postulates of statistical mechanics [18] become valid and one
can depict the system under the premises of equilibrium (but still quantum)
statistical mechanics. The system has to be in a sufficiently high temperature
when the spin-statistics can be adequately represented by the Maxwell-
Boltzmann distribution (particle becomes distinguishable) for it to be
aptly called classical.

%\end

\vfill
\eject

\centerline{\bf 2. Quantum and Thermal Fluctuations in an Open System}

\vskip 0.6cm

\noindent{\bf 2.1 Influence Functional for a Brownian Particle in a
Harmonic Oscillator Bath}

\vskip 0.5cm

Consider a Brownian particle with mass $M$ and natural frequency $\Omega_0$.
Its environment is modeled by a set of harmonic oscillators with mass $m_n$
and natural frequency $\omega_n$. The Brownian particle is coupled linearly
to the $n$th oscillator with strength $C_n$. The action of the combined
system and environment is

$$
{\eqalign{
S[x,q]
& = S[x]+S_b[q]+S_{int}[x,q] \cr
& = \int\limits_0^tds\Biggl[
    {1\over 2}M\Bigl(\dot x^2-\Omega_0^2 x^2\Bigr)
  + \sum_n\Bigl\{{1\over 2}m_n\dot q_n^2
  - {1\over 2}m_n\omega^2_nq_n^2 \Bigr\}
  + \sum_n\Bigl\{-C_nxq_n\Bigr\}\Biggr]\cr}}                \eqno(2.1.1)
$$

\noindent where $x$ and $q_n$ are the coordinates of the Brownian particle
and the $n$-th bath oscillators respectively. We are interested in how the
environment affects the system in some averaged way. The quantity containing
this information is the reduced density matrix of the system $\rho_r(x,x',t)$
obtained from the full density operator of the system and environment
$\rho(x,q;x',q';t)$ by tracing out the environmental degrees of freedom
$(q_n,q'_n)$

$$
\rho_r(x,x',t)
=\prod_n\int\limits_{-\infty}^{+\infty}dq_n
 \int\limits_{-\infty}^{+\infty}dq'_n~
 \rho(x,q;x',q',t)\delta(q_n-q'_n)                         \eqno(2.1.2)
$$

\noindent The reduced density matrix evolves under the action of the
evolutionary operator $J_r$ in the following way:

$$
\rho_r(x,x',t)
=\int\limits_{-\infty}^{+\infty}dx_i
 \int\limits_{-\infty}^{+\infty}dx'_i~
 J_r(x,x',t~|~x_i,x'_i,0)~\rho_r(x_i,x'_i,0~)              \eqno(2.1.3)
$$

\noindent In general, this is a very complicated expression since
$J_r$ depends on the initial state. If we assume that
at a given time $t=0$ the system and the environment are uncorrelated,
i.e. that

$$
\hat\rho(t=0)
=\hat\rho_s(t=0)\times\hat\rho_e(t=0),                      \eqno(2.1.4)
$$

\noindent then the evolution operator for the reduced density matrix can
be written as

$$
J_r(x_f,x'_f,t~|~x_i,x'_i,0)
=\int\limits_{x_i}^{x_f}Dx
 \int\limits_{x'_i}^{x'_f}Dx'~
 \exp{i\over\hbar}\Bigl\{S[x]-S[x']\Bigr\}~F[x,x']
                                                            \eqno(2.1.5)
$$

\noindent where $F[x,x']$ is the Feynman-Vernon influence functional [12].
If the environment is initially in thermal equilibrium at temperature $T$,
for the problem described by (2.1.1), the influence functional can be
computed exactly. The result is well known [12-15]:

$$
\eqalign{
F[x,x']=\exp\biggl\{
& -{i\over\hbar}\int\limits_0^tds_1\int\limits_0^{s_1}ds_2
   \Bigl[x(s_1)-x'(s_1)\Bigr]\eta(s_1-s_2)
   \Bigl[x(s_2)+x'(s_2)\Bigr] \cr
& -{1\over\hbar}\int\limits_0^tds_1\int\limits_0^{s_1}ds_2
   \Bigl[x(s_1)-x'(s_1)\Bigr]\nu(s_1-s_2)
   \Bigl[x(s_2)-x'(s_2)\Bigr]\biggl\} \cr}
                                                           \eqno(2.1.6)
$$

\noindent The non-local kernels $\eta(s)$ and $\nu(s)$ are defined as

$$
\eqalignno{
\nu(s)&=\int\limits_0^{+\infty}{d\omega\over\pi}~
        I(\omega)\coth{\hbar\omega \over 2kT}\cos\omega s  &(2.1.7a)\cr
\noalign{\hbox{and}}
\eta(s)&={d\over ds}~\gamma(s)                             &(2.1.7b)\cr
\noalign{\hbox{where}}
\gamma(s)&=\int\limits_0^{+\infty}{d\omega\over\pi}
           ~{I(\omega)\over\omega}~\cos\omega s            &(2.1.7c)\cr}
$$

\noindent Here $I(\omega)$ is the spectral density function of the
environment,

$$
I(\omega)= \sum\limits_n{\delta(\omega -\omega_n)}
           {\pi C^2_n\over 2m_n\omega_n}                   \eqno(2.1.8)
$$

\noindent The real and imaginary parts of the exponent in the influence
functional $F[x,x']$ are understood as responsible for dissipation and
noise respectively, thus the names dissipation and noise kernels are given
to $\eta(s)$ and $\nu(s)$. The most general environment would thus engender
nonlocal dissipation and colored noise. (We refer the reader to Ref [1, 15] for
a discussion of the fluctuation-dissipation relation and the time scales  of
the relevant processes.)

An example of the spectral density is given by [13]

$$
I(\omega)
=M\gamma_0 \omega \Bigl({\omega\over \tilde\omega}\Bigr)^s
e^{-{\omega^2\over\Lambda^2}}                             \eqno(2.1.9)
$$

\noindent where $\tilde\omega$ is a frequency scale usually taken to be
the cut-off frequency $\Lambda$. The  environment with this kind of spectral
density is classified as

$$
\eqalignno{
ohmic
& ~~~ if ~~ s=1 ~~~~~                  &(2.1.10a) \cr
supra-ohmic
& ~~~ if ~~ s>1 ~~~~~                  &(2.1.10b) \cr
subohmic
& ~~~ if ~~ s<1 ~~~~~                  &(2.1.10c) \cr }
$$

The propagator (2.1.5) has been calculated before [12-15]. (We use the
notation of Ref. 15)

$$
\eqalign{
J_r(x_f,x'_f,t~|~x_i,x'_i,0)
& = Z_0(t) \exp{i\over\hbar} \Biggl\{
    \Bigl[\dot u_1(0)X_i+\dot u_2(0)X_f\Bigr]Y_i                \cr
&  -\Bigl[\dot u_1(t)X_i+\dot u_2(t)X_f\Bigr]Y_f
   +ia_{11}(t)Y_i^2                                             \cr
&  +i\Bigl[a_{12}(t)+a_{21}(t)\Bigr]Y_iY_f
   +ia_{22}(t)Y_f^2  \Biggr\}                                   \cr }
                                                        \eqno(2.1.11)
$$

\noindent Here we have introduced

$$
\cases{
\eqalign{
&X_i={1\over 2}(x_i+x'_i)     \cr
&Y_i=x'_i-x_i                 \cr} }
                                                       \eqno(2.1.12a)
$$

\noindent and

$$
\cases{
\eqalign{
&X_f={1\over 2}(x_f+x'_f)     \cr
&Y_f=x'_f-x_f                 \cr} }
                                                       \eqno(2.1.12b)
$$

\noindent The elementary functions $u_i(s), i=1,2$ in (2.1.11) satisify
the following integro-differential equation

$$
  {d^2 u_i(s)\over ds^2}
  +2\int\limits_0^sds'\eta((s-s')u_i(s')
  +\Omega^2_0u_i(s)=0
                                                      \eqno(2.1.13)
$$

\noindent with boundary conditions

$$
\cases{
\eqalign{
& u_1(0)=~1    \cr
& u_1(0)=~0    \cr  } }
\qquad\hbox{and}\qquad
\cases{
\eqalign{
& u_2(0)=~0    \cr
& u_2(0)=~1    \cr  } }
                                                      \eqno(2.1.14)
$$

\noindent The coefficient functions $ a_{ij}(t), ~i,j=1,2 $ are defined by

$$
a_{ij}(t)
=\int\limits_0^tds_1\int\limits_0^{s_1}ds_2
 v_i(s_1)\nu(s_1-s_2)v_j(s_2)
                                                      \eqno(2.1.15)
$$

\noindent where

$$
v_1(s)\equiv u_2(t-s)
                                                      \eqno(2.1.16a)
$$

\noindent and

$$
v_2(s)\equiv u_1(t-s)
                                                      \eqno(2.1.16b)
$$

\noindent It is clear that the functions $u_i(t)$ and $a_{ij}(t)$ only
depend on the kernels $\eta(s)$ and $\nu(s)$.

\vfill
\eject

\noindent{\bf 2.2 Reduced Density Matrix For A Gaussian Wavepacket }

\vskip 0.25cm

We now consider a Brownian harmonic oscillator with an initial wave function
(Calderia and Leggett 1985, Unruh and Zurek 1987 in [2],
Hu, Paz and Zhang 1992 in [15], Paz, Habib and Zurek 1992 in [16])
$$
\psi(x,0)=
\sqrt{N_0}~\exp\Bigl\{-{(x-x_0)^2\over 4\sigma^2}
+{i\over\hbar}p_0 x \Bigr\}
                                                      \eqno(2.2.1)
$$

\noindent where $\sigma$ is the initial spread, and $x_0$ and $p_0$ are the
averaged initial position and momentum of the Gaussian wave packet.
The initial reduced density matrix is given by

$$
\eqalign{
\rho_r(x_i,x_i',0)
&=N_0\exp\biggl\{
-{(x_i-x_0)^2\over 4\sigma^2}
-{i\over\hbar} p_0 x_i
-{(x'_i-x_0)^2\over 4\sigma^2}
+{i\over\hbar} p_0 x'_i \biggl\}    \cr
&=\tilde N_0\exp\biggl\{
-{X_i^2\over 2\sigma^2}
-{Y_i^2\over 8\sigma^2}
+{x_0\over\sigma^2} X
+{i\over\hbar} p_0 Y \biggl\}         \cr }
                                                        \eqno(2.2.2)
$$

\noindent where

$$
\tilde N_0 = N_0 \exp\Bigl\{ -{x_0^2\over 2\sigma^2} \Bigl\}
                                                        \eqno(2.2.3)
$$

\noindent Substituting this and the propogator (2.1.11) into (2.1.3),
one would get

$$
\eqalign{
\rho_r(x_f,x'_f,t)
& = Z_0 \tilde N_0 \exp{i\over\hbar}
   \Bigl\{-\dot u_2(t)X_fY_f+ia_{22}(t)Y_f^2\Bigr\} \times
   \int\limits_{-\infty}^{+\infty}dX
   \int\limits_{-\infty}^{+\infty}dY                          \cr
&  \times \exp\Biggl\{-{1\over 2}
      \pmatrix{ X_i \cr Y_i \cr }^T
      \pmatrix{ H_{11}(t) & H_{12}(t) \cr
                H_{21}(t) & H_{22}(t) \cr }
      \pmatrix{ X_i \cr Y_i \cr }
    + \pmatrix{ b_1(t) \cr b_2(t) \cr}^T
      \pmatrix{ X_i    \cr Y_i    \cr}   \Biggr\}             \cr }
                                                          \eqno(2.2.4)
$$

\noindent where

$$
\eqalignno{
b_1(t)
&  = -{i\over\hbar}\dot u_1(t) Y_f
   + {x_0\over \sigma^2}                                 &(2.2.5a)\cr
b_2(t)
&  =  {i\over\hbar}\dot u_2(0) X_f
   - {1\over\hbar}\bigl[a_{12}(t)+a_{21}(t)\bigr] Y_f
   + {i\over\hbar} p_0                                   &(2.2.5b)\cr }
$$

\noindent and the matrix elements of ${\bf H(t)}$ are given by

$$
\eqalignno{
H_{11}(t) & = {1\over\sigma^2}              &(2.2.6)a\cr
H_{12}(t) & = H_{21}(t)
            = - {i\over\hbar}\dot u_1(0)    &(2.2.6b)\cr
H_{22}(t) & = {1\over 4\sigma^2}
            + {2\over\hbar} a_{22}(t)       &(2.2.6c)\cr}
$$

\noindent Performing the Gaussian integrals over $X$ and $Y$ in (2.2.4),
one gets

$$
\rho_r(x_f, x'_f, t)
=\tilde Z_0(t)\exp\Biggl\{
 -{1\over 2}\pmatrix{ X_f \cr Y_f \cr }^T
            \pmatrix{ Q_{11}(t) & Q_{12}(t) \cr
                      Q_{21}(t) & Q_{22}(t) \cr }
            \pmatrix{ X_f \cr Y_f \cr }
          + \pmatrix{ B_1(t) \cr B_2(t) \cr}^T
            \pmatrix{ X_f    \cr Y_f    \cr}   \Biggr\}
                                                           \eqno(2.2.7)
$$

\noindent where

$$
B_1(t)
= -{i\dot u_2(0) H_{12}\over \hbar\sigma^2 \det {\bf H} (t)} x_0
  -{\dot u_2(0)H_{11}(t)\over \hbar^2 \det {\bf H} (t)} p_0 ,
                                                          \eqno(2.2.8a)
$$

$$
\eqalign{
B_2(t)
&={1\over\hbar\det {\bf H} (t)}\Bigl\{
  -iH_{22}(t)+H_{12}(t)\bigl[a_{12}(t)
  +a_{21}(t)\bigr]\Bigl\} x_0                              \cr
& -{1\over\sigma^2\hbar\det {\bf H} (t)} \Bigl\{
   \dot u_1(t) H_{12}(t)
  +iH_{11}(t)\bigl[a_{12}(t)+a_{21}(t)\bigr] \Bigl\} p_0   \cr}
                                                           \eqno(2.2.8b)
$$

\noindent The prefactor

$$
\tilde Z_0(t)
=Z_0(t)\tilde N_0 {\pi\over\sqrt{\det{H}}} \exp\biggl\{
 {H_{22}(t)\over 2\sigma^2\det {\bf H} (t)} x_0^2
-{iH_{12}(t)\over \hbar\sigma^2\det {\bf H} (t)} x_0 p_0
-{H_{11}\over 2\hbar^2\det {\bf H} (t)} p_0^2 \biggr\}
                                                          \eqno(2.2.9)
$$

\noindent depends only on the parameters $x_0$, $p_0$ and time. The
matrix elements of ${\bf Q}(t)$ are given by

$$
Q_{11}(t)
= {\bigl[\dot u_2(0)\bigr]^2 \over \hbar^2 \sigma^2 \det {\bf H} (t) }
                                                            \eqno(2.2.10a)
$$

$$
Q_{12}(t) = Q_{21}(t)
= {i\over\hbar} \dot u_2(t)
+ {i\over\hbar^2} {\dot u_2(0)\over\det {\bf H} (t)}
\biggl\{{1\over\sigma^2}\Bigl[a_{12}(t)+a_{21}(t)\Bigr]
- {1\over\hbar}\dot u_1(0) \dot u_1(t) \biggr\}
                                                            \eqno(2.2.10b)
$$

$$
\eqalign{
Q_{22}(t)
& = {2\over\hbar} a_{22}(t)
  + {1\over \hbar^2\det {\bf H} (t)} \biggl\{
    \Bigl[ {1\over 4\sigma^2}
  + {2\over\hbar} a_{11}(t) \Bigr]
    \bigl[\dot u_1(t)\bigr]^2                      \cr
& + {2\over\hbar} \Bigr[a_{12}(t)+a_{21}(t)\Bigr]
    \dot u_1(0) \dot u_1(t)
  - {1\over\sigma^2} \Bigl[a_{12}(t)+a_{21}(t)\Bigr]^2 \biggr\}  \cr }
                                                           \eqno(2.2.10c)
$$

\vskip 1.0cm

\noindent{\bf 2.3 Finite Temperature Quantum Uncertainty Function}

\vskip 0.5cm

To calculate the averages of observables, it is convenient to use the
Wigner function associated with the reduced density matrix $\rho_r$
defined as

$$
W_r(X,p,t)
= \int\limits_{-\infty}^{+\infty} dY ~ e^{{i\over \hbar} pY} ~
  \rho_r (X-{1\over 2}Y, X+{1\over 2}Y, t),                  \eqno(2.3.1a)
$$

\noindent with an inverse relation given by

$$
\rho_r (X-{1\over 2}Y, X+{1\over 2}Y, t)
= \int\limits_{-\infty}^{+\infty} {dp\over 2\pi\hbar} ~
  e^{-{i\over\hbar} pY}~ W_r (X, p, t)                       \eqno(2.3.1b)
$$

\noindent Applying the above Wigner transform to the reduced density
matrix (2.2.7), we get

$$
\eqalign{
W_r(X,p,t)
&=\int\limits_{-\infty}^{+\infty} dY ~
  e^{{i\over\hbar}pY} ~ \tilde Z_0(t)             \cr
& \times \exp\Biggl\{-{1\over 2}
  \pmatrix{ X_f \cr Y_f \cr }^T
  \pmatrix{ Q_{11}(t) & Q_{12}(t) \cr
            Q_{21}(t) & Q_{22}(t) \cr }
  \pmatrix{ X_f \cr Y_f \cr }
 +\pmatrix{ B_1(t) \cr B_2(t) \cr}^T
  \pmatrix{ X_f    \cr Y_f    \cr }   \Biggr\} \cr }
                                                            \eqno(2.3.2)
$$

\noindent where the superscript T denotes the transpose of the matrix.
The quantum statistical averages of an observable of the system, e.g.,
$x^n$, or $p^n$ with respect to the reduced density matrix $\rho_r(t)$
are given by

$$
< x^n>_T
= \int\limits_{-\infty}^{+\infty} dx
  \int\limits_{-\infty}^{+\infty} {dp\over 2\pi\hbar}~
  x^n ~ W_r(X, p, t)
                                                            \eqno(2.3.3)
$$

\noindent and

$$
< p^n>_T
= \int\limits_{-\infty}^{+\infty} dx
  \int\limits_{-\infty}^{+\infty} {dp\over 2\pi\hbar}~
    p^n ~ W_r(X, p, t)
                                                             \eqno(2.3.4)
$$

\noindent where the subscript $T$ indicates that the environment is at
temperature $T$. Obviously the averages have both quantum and thermal
contributions. It is easy to show that

$$
< x >_T = {B_1(t)\over Q_{11}(t)}
                                                             \eqno(2.3.5)
$$

\noindent and

$$
< x^2 >_T
= {1\over Q_{11}(t)} + \Bigl[ < x >_T \Bigl]^2
                                                             \eqno(2.3.6)
$$

\noindent thus

$$
(\Delta x)^2 =~<x^2>_T-<x>_T^2~=~{1\over Q_{11}(t)}
                                                             \eqno(2.3.7)
$$

\noindent Similarly,

$$
< p >_T
=i\hbar^2\Bigl[{\det {\bf Q} (t)\over Q_{11}(t)}\Bigl]^{1\over 2}~
 \Bigl[B_2(t)-B_1(t){Q_{12}(t)\over Q_{11}(t)}\Bigr]
                                                             \eqno(2.3.8)
$$

\noindent and

$$
< p^2 >_T
= \hbar^2 {\det {\bf Q}\over Q_{11}(t)}
+ \Bigl[ < p >_T \Bigr]^2
                                                             \eqno(2.3.9)
$$

\noindent thus

$$
(\Delta p)_T^2 =~<p^2>_T-<p>_T^2~=~\hbar^2{\det {\bf Q}\over Q_{11}(t)}
                                                             \eqno(2.3.10)
$$

\noindent The finite temperature quantum uncertainty function defined as

$$
U_T(t)=(\Delta x)_T^2 ~ (\Delta p)_T^2
                                                             \eqno(2.3.11)
$$

\noindent follows:

$$
\eqalign{
U_T(t)
&=\hbar^2{\det {\bf Q} (t)\over [Q_{11}(t)]^2}                       \cr
&=\hbar^2\Biggl\{ {Q_{22}(t)\over Q_{11}(t)}
        -\Bigl[   {Q_{12}(t)\over Q_{11}(t)} \Bigr]^2 \Biggr\} \cr }
                                                             \eqno(2.3.12)
$$

\noindent Using (2.2.10a-c), one can rewrite

$$
{Q_{12}(t)\over Q_{11}(t)} = if_0(t) + if_1(t)
                                                             \eqno(2.3.13)
$$

\noindent and

$$
{Q_{22}(t)\over Q_{11}(t)} = g_0(t) + g_1(t) + g_2(t)
                                                             \eqno(2.3.14)
$$

\noindent where the functions $f_n, g_n$ denote terms of order $n$ in
temperature $T$. Explicitly  they are given by:
$$
f_0(t)
= {\sigma^2\over\hbar}
  {\dot u_2(t)\over \bigl[\dot u_2(0)\bigr]^2}
   \biggl\{ {\hbar^2\over 4\sigma^4}
+ {\dot u_1(0)\over\dot u_2(t)}
   \Bigl[\dot u_1(0)\dot u_2(t)
-        \dot u_1(t)\dot u_2(0) \Bigr] \biggr\},
                                                            \eqno(2.3.15)
$$

$$
f_1(t)
= {1\over \dot u_2(0)} \biggl\{
  2{\dot u_2(t)\over\dot u_2(0)} a_{11}(t)
+ a_{12}(t)+a_{21}(t) \biggl\}                ,
                                                            \eqno(2.3.16)
$$

\noindent and

$$
g_0(t)
={1\over 4}\Bigl[{\dot u_1(t)\over\dot u_2(0)}\Bigr]^2,
                                                            \eqno(2.3.17)
$$

$$
\eqalign{
g_1(t)
&= {2\sigma^2\over \hbar \bigl[\dot u_2(0)\bigr]^2}
   \biggl\{ \bigl[\dot u_1(t)\bigr]^2a_{11}(t)
   +\dot u_1(0) \dot u_1(t)                             \cr
&  \times \Bigl[a_{12}(t)+a_{21}(t)\Bigr]
   +\Bigl[ {\hbar^2\over 4\sigma^4}
   +\bigl[\dot u_1(0)\bigr]^2\Bigr] a_{22}(t) \biggr\},  \cr }
                                                           \eqno(2.3.18)
$$

$$
g_2(t)
={1\over\big[\dot u_2(0)\bigr]^2}
\biggl\{4a_{11}(t)a_{22}(t)-
\bigl[a_{12}(t)+a_{21}(t)\bigr]^2 \biggr\}.
                                                           \eqno(2.3.19)
$$

\noindent The finite temperature uncertainty function is then formally given by

$$
U_T(t) = \Bigl[ g_0(t) + g_1(t) + g_2(t) \Bigr]
       + \Bigl[ f_0(t) + f_1(t) \Bigr]^2
                                                           \eqno(2.3.20)
$$

\vfill
\eject

%\end
%From:	UMDHEP::ZHANG        15-JUN-1993 23:38:39.06

%\magnification=\magstep1
%\baselineskip=24truept
%\pageno=1

\centerline{\bf 3. Uncertainty Principle and Squeezed States}

\vskip 0.5cm

\noindent{\bf 3.1 Finite Temperature Ohmic Environment}

\vskip 0.5cm

We now proceed to analyze the finite temperature uncertainty
function for an  ohmic environment. In the cases of high temperature and
weak couplings, one can obtain relatively simple
analytic expressions
%We can use these results to discuss a few issues:
%1) the relative importance of quantum and thermal fluctuations
%in the uncertainty relation,
%2) its relation to the decoherence time, and
%3) the transition of the
%system from a quantum nonequilibrium condition to that describable by
%classical equilibrium statistical mechanics. Throughout we shall
%discuss these issues as manifested in
for the uncertainty relation for a general squeezed states.
Results for the (unsqueezed) coherent state or the {\it minimal
uncertainty state} were obtained by Hu and Zhang [5] earlier.
For simplicity, we set the mass of the Brownian harmonic oscillator $M=1$.
The spectral density for ohmic dissipation is

$$
I(\omega) = \gamma_0 \omega ~ e^{-{\omega^2\over\Lambda^2}}
                                                         \eqno(3.1.1)
$$

\noindent With this, the dissipation kernel becomes

$$
\eqalign{
\gamma(s)
& = \gamma_0\int\limits_0^{+\infty}{d\omega\over\pi}
    \cos\omega s ~ e^{-{\omega^2\over\Lambda^2}}        \cr
& = \gamma_0 {\Lambda\over 2\pi^{1/2}}
    e^{-{1\over 4}\Lambda^2 s^2}                        \cr }
                                                        \eqno(3.1.2)
$$

\noindent If we assume that the cutoff frequency $\Lambda$ is very large,
then

$$
\gamma(s) \simeq \gamma_0 \delta(s)
                                                         \eqno(3.1.3)
$$

\noindent so the damping becomes local. However, the noise kernel

$$
\nu(s)
=\gamma_0\int\limits_0^{+\infty}{d\omega\over\pi}~
  \omega\coth{\hbar\omega \over 2kT}~
  e^{-{\omega^2\over\Lambda^2}}~\cos\omega s
                                                         \eqno(3.1.4)
$$

\noindent is generally nonlocal except at very high temperatures,

$$
{\hbar\Lambda\over kT} << 1
                                                         \eqno(3.1.5)
$$

\noindent whence

$$
\nu(s)\simeq {2kT\over\hbar} \gamma_0 \delta(s)
                                                         \eqno(3.1.6)
$$

\noindent corresponding to white noise.

For local dissipation, equation (2.1.13) becomes

$$
{d^2 u_i(s)\over ds^2} + \gamma_0 {d u_i(s)\over ds}
+ \Bigl[\Omega^2_0 - 2 \gamma_0 \delta(0) \Bigr] u_i(s)
= - 2\delta(s) u_i(0)
                                                         \eqno(3.1.7)
$$

\noindent The boundary term on the right side of equation (3.1.7) can be
neglected since it gives no contribution to the solutions with the boundary
conditions (2.1.14). We also can redefine the natural frequency by

$$
\Bigl\{ \Omega^2_0-2 \gamma_0 \delta(0) \Bigr\} \to \Omega^2_0,
                                                         \eqno(3.1.8)
$$

\noindent then equation (3.1.7) becomes the familiar equation of motion
for a damped harmonic oscillator. Let us define a damping parameter

$$
\alpha \equiv {\gamma_0 \over 2\Omega}
                                                          \eqno(3.1.9)
$$
and the effective frequency [15]
$$
\Omega=(\Omega_0^2-{1\over 4}\gamma_0^2)^{1/2}
                                                          \eqno(3.1.10)
$$
Then in the case of underdamping, $\alpha < 1$
the elementary functions simplify to

$$
\eqalignno{
u_1(s) & = {\sin\Omega(t-s)\over\sin\Omega t}~
           e^{-{1\over 2}\gamma_0 s}
                                                          &(3.1.11a)\cr
\noalign{\hbox{and}}
u_2(s) & = {\sin\Omega s\over\sin\Omega t}~
           e^{{1\over 2}\gamma_0 (t-s)}
                                                          &(3.1.11b)\cr }
$$
The derivatives of these elementary functions are
$$
\eqalignno{
\dot u_1(0)
  & = -\Omega{\cos\Omega t\over\sin\Omega t}
      -{1\over 2}\gamma_0                                  &(3.1.12a)\cr
\dot u_1(t)
  & = -{\Omega\over\sin\Omega t}~
      e^{-{1\over 2}\gamma_0 t}                            &(3.1.12b)\cr
\noalign{\hbox{and}}
\dot u_2(0)
  & = {\Omega\over\sin\Omega t}~
      e^{{1\over 2}\gamma_0 t}                             &(3.1.12c)\cr
\dot u_2(t)
  & = \Omega{\cos\Omega t\over\sin\Omega_0t}
      -{1\over 2}\gamma_0                                  &(3.1.12d)\cr}
$$

\noindent By using Eq. (3.1.12a-d), it can be shown that

$$
f_0(t)
= {\Omega_0\over\Omega} \biggl\{
  {{1-\delta^2}\over 4\delta}\sin 2\Omega t
- \alpha
  {{1+\delta^2}\over 2\delta}\sin^2\Omega t \biggr\}
  e^{-\gamma_0 t}
                                                           \eqno(3.1.13)
$$

\noindent and

$$
g_0(t) = {1\over 4} e^{-2\gamma_0 t}
                                                           \eqno(3.1.14)
$$

\noindent where we have introduced a squeeze parameter

$$
\delta={2\Omega_0\sigma^2 \over \hbar}
                                                           \eqno(3.1.15)
$$

\noindent which measures the spread in the initial Gaussian wavepacket.
($\delta =1 $ corresponds to a coherent state or  minimal-uncertainty
state.)

It can be shown that (see Appendix)

$$
f_1(t)={2\over\Omega}
      \Bigl\{-\alpha_0[ss]+[sc]\Bigr\}
                                                           \eqno(3.1.16)
$$

$$
\eqalign{
g_1(t)
& = \delta {\Omega_0\over 2\Omega^2} \biggl\{
    \Bigl[1+{1\over\delta^2}\Bigl]
   -{\gamma_0\Omega\over\Omega_0^2} \sin 2\Omega t
   -\Bigl[1-{1\over\delta^2}
   -{\gamma_0^2\over 2\Omega_0^2}\Bigr] \cos 2\Omega t
    \biggr\} [ss]                                                    \cr
&  +{\delta\over 2\Omega_0^2} \biggl\{
    \Bigl[1-{1\over\delta^2}
   -2{\gamma_0^2\over 2\Omega_0^2}\Bigr] \sin 2\Omega t
   -{\gamma_0\Omega\over\Omega_0^2} \cos 2\Omega t
    \biggr\} [sc]                                                    \cr
&  +{\delta\over 4\Omega_0^2} \biggl\{
    \Bigl[1+{1\over\delta}\Bigl]
   +{\gamma_0\Omega\over\Omega_0^2} \sin 2\Omega t
   +\Bigl[1-{1\over\delta^2}
   -{\gamma_0^2\over 2\Omega_0^2}\Bigr] \cos 2\Omega t
    \biggr\} [cc]                                                     \cr }
                                                           \eqno(3.1.17)
$$

\noindent and

$$
g_2(t)={4\over\Omega^2}\biggl\{[ss][cc]-[sc]^2\biggr\}
                                                           \eqno(3.1.18)
$$

\noindent where the functions $[ss]$, $[sc]$ and $[cc]$ are defined in (A.1)

\vskip 0.5cm

\noindent{\bf 3.2 Zero Coupling Limit }

\vskip 0.5cm

Let us first examine the simplest case of zero-coupling. It corresponds
to an isolated harmonic oscillator taken as a closed quantum system.
We expect to
recover the familiar results in quantum mechanics.

Assuming $ \gamma_0 = 0$, then $ \gamma(s) = 0$ and  $ \nu(s) = 0$. Therefore,

$$
f_1(s) = g_1(t) = g_2(t) = 0
                                                         \eqno(3.2.1)
$$

\noindent and

$$
\Omega=\Omega_0
                                                         \eqno(3.2.2)
$$

\noindent From Eq. (3.1.14) and (3.1.15) one gets

$$
f_0(t) = {1-\delta^2 \over 4\delta}\sin 2\Omega_0t
                                                         \eqno(3.2.3)
$$

\noindent and

$$
g_0(t) = {1\over 4}
                                                         \eqno(3.2.4)
$$

\noindent We find the quantum uncertainty function for an initial
squeezed states to be

$$
U_T(t)
= {\hbar^2 \over 4}
\Bigl\{ 1 + { (1-\delta^2)^2 \over 4\delta^2 }
\sin^22\Omega_0t \Bigr\}
\ge {\hbar^2 \over 4}
                                                          \eqno(3.2.5)
$$
The time-dependent term is the result of quantum dispersion.
For the (unsqueezed) coherent state,
$$
\delta = 1
                                                          \eqno(3.2.6)
$$
\noindent we recover the Heisenberg uncertainty relation
$$
U_T(t) = {\hbar^2  \over 4}                             \eqno(3.2.7)
$$
\noindent With this we can also understand why $\delta =1$ is called a
{\it minimun-uncertainty state}.

\vskip 0.5cm

\noindent{\bf 3.3 High Temperature Limit }

\vskip 0.25cm

As one can see from (3.1.3) and (3.1.6), at high temperatures both the noise
kernel and the dissipation kernel for ohmic dissipation become local.
In this limit the functions $f$ and $g$ simplify to (See Appendix)

$$
f_1(t) = \alpha \tau_0 \sin^2\Omega t  e^{-\gamma_0 t}
                                                          \eqno(3.3.1)
$$

$$
g_1(t)
= \tau_0 \biggl\{ {{1+\delta^2}\over 4\delta} \Bigl[1-e^{-\gamma_0 t}\Bigr]
+ {{1+\delta^2}\over 2\delta}  \alpha^2 \sin^2\Omega t
 - {{1-\delta^2}\over 4\delta} \alpha\sin 2\Omega t \biggr\}
   e^{-\gamma_0 t}
                                                         \eqno(3.3.2a)
$$

\noindent and

$$
g_2(t)
= \tau_0^2 \biggl\{ {1\over 4} -\Bigl[{1\over 2}+\alpha^2
  \sin^2\Omega t\Bigr] e^{-\gamma_0 t}
+ {1\over 4} e^{-2\gamma_0 t}  \biggr\}
                                                          \eqno(3.3.2b)
$$

\noindent where we have introduced the dimensionless parameter
$$
\tau_0 \equiv {{2k_BT} \over {\hbar \Omega_0}} \equiv {1 \over \epsilon_0}
                                                           \eqno(3.3.3)
$$
We will also use $\epsilon (\omega)$ and $\epsilon$ to denote the
same quantities as $\epsilon_0$, but  with
$\omega $ and $\Omega$ replacing $\Omega_0$ in (3.3.3) respectively.
The uncertainty function for ohmic dissipation
at the high temperature limit is  then given by

$$
\eqalign{
{1\over\hbar^2}U_T(t)
& = {1 \over 4}\biggl\{e^{-\gamma_0 t}
    + \tau_0 \Bigl[1-e^{-\gamma_0 t}\Bigr] \biggl\}^2             \cr
& + \tau_0 {(1-\delta)^2\over 4\delta}
    \Bigl[1-e^{-\gamma_0 t}\Bigr] e^{-\gamma_0 t}               \cr
& - \tau_0 \alpha
     \biggl\{ {{1-\delta^2}\over 4\delta} \sin 2\Omega t
    +\alpha \Bigl[ \tau_0
    -{{1+\delta^2} \over 2\delta} \Bigr]
     \sin^2\Omega t \biggr\} e^{-\gamma_0 t}                  \cr
& + {\Omega_0^2\over\Omega^2} \biggl\{
     {{1-\delta^2}\over 4\delta} \sin 2\Omega t
  +  \alpha \Bigl[ \tau_0 -{{1+\delta^2} \over 2\delta} \Bigr]
     \sin^2\Omega t \biggr\}^2 e^{-2\gamma_0 t}               \cr}
                                                        \eqno(3.3.4)
$$

\noindent This is the first main result of this paper.

Alternatively, since the noise kernel is also a Dirac delta function, from
(2.1.15), we have

$$
a_{ij}(t)
\simeq {kT\over\hbar}\gamma_0 \int\limits_0^t ds ~v_i(s)v_j(s)
                                                         \eqno(3.3.5)
$$

\noindent They are easily computed to be

$$
a_{11}(t)
\simeq {kT\over\hbar}
 {\Omega^2\over 2\Omega_0^2\sin^2\Omega t} e^{\gamma_0 t}
 \biggl\{ 1 - \Bigl[ {\Omega_0^2\over\Omega^2}
 -\alpha^2\cos 2\Omega t
 +\alpha\sin 2\Omega t \Bigr]
  e^{-\gamma_0t} \biggr\}
                                                          \eqno(3.3.6a)
$$

$$
a_{12}(t)
\simeq {kT\over\hbar}
 {\Omega^2\over 2\Omega_0^2\sin^2\Omega t}
  e^{{1\over 2}\gamma_0t} \biggl\{
  -\Bigl[ \cos\Omega t-\alpha\sin\Omega t \Bigr]
  +\Bigl[ \cos\Omega t+\alpha\sin\Omega t \Bigr]
  e^{-\gamma_0t} \biggr\}
                                                           \eqno(3.3.6b)
$$

$$
a_{22}(t)
\simeq {kT\over\hbar}
 {\Omega^2\over 2\Omega_0^2\sin^2\Omega t}\biggl\{
  \Bigl[ {\Omega_0^2\over\Omega^2}
 -\alpha^2\cos 2\Omega t
 -\alpha\sin 2\Omega t \Bigr]
 -e^{-\gamma_0t} \biggr\}
                                                           \eqno(3.3.6c)
$$

\noindent Substituting these and (3.1.12a-d) into (2.3.16) and (2.3.18-20),
we can also arrive at Eq.(3.3.2-4).

We see that
there are two factors at play here: time and temperature. Time is measured
in units of the relaxation time proportional to $t_{rel}=\gamma_0^{-1}$, and
temperature is measured with reference to the ground state energy
$\hbar \Omega_0 /2$ of the system. There are also two parameters involved,
$\alpha$ and $\delta$.
Let us now take a closer look at this high temperature uncertainty function
under different conditions:

\vskip 0.2cm

\noindent a) At $t=0$, we find

$$
U_T(0) = {\hbar^2 \over 4}
                                                        \eqno(3.3.7)
$$

\noindent At  $t >> \gamma_0^{-1}$, we find

$$
U_T(t) \simeq \Bigl({kT\over\Omega_0}\Bigr)^2
                                                        \eqno(3.3.8)
$$

\noindent These comply with the  expected results from quantum mechanics
and classical statistical mechanics as stated in the Introduction.

\vskip 0.2cm

\noindent b) At short times $t << \gamma_0^{-1}$,
we can expand the uncertainty function (3.3.4) as

$$
U_T(t) \simeq {\hbar^2\over 4}[ 1 +  2(\tau_0 \delta - 1) \gamma_0 t
        + O(t^2)]                                         \eqno(3.3.9)
$$

\noindent From this we can identify the time when thermal fluctuations
overtake quantum fluctuations, i.e.,

$$
t_1 \simeq {\hbar\Omega_0\over 4kT\gamma_0\delta}
   ~~~(high~~temperature)                                  \eqno(3.3.10)
$$

\noindent which is seen to depend on $\delta$, the squeeze parameter of
the initial state. This time is identical to the decoherence time derived
in  [2] [15, 16]. That the transition time from quantum to thermal
dominance in the uncertainty relation is related to the
decoherence time scale was first noted by the present authors in [5].

\vskip 0.2cm

\noindent c) For a minimum uncertainty initial state ($\delta=1$),

$$
\eqalign{
{1\over\hbar^2}U_T(t)
& ={1\over 4} \biggl\{ e^{-\gamma_0 t} + \tau_0
     \Bigl[1-e^{-\gamma_0 t}\Bigr] \biggl\}^2                    \cr
&  - \tau_0 \alpha^2
     (\tau_0 -1)
     \sin^2\Omega t e^{-\gamma_0 t}                            \cr
& + {\Omega_0^2\over\Omega^2} \alpha^2
     (\tau_0 -1)^2
     \sin^4\Omega t e^{-2\gamma_0 t}                           \cr}
                                                        \eqno(3.3.11)
$$

\vskip 0.2cm

\noindent d) For weak damping, one can neglect all $\gamma_0/\Omega_0$
terms and find

$$
\eqalign{
{1\over\hbar^2}U_T(t)
& = {1 \over 4}\Biggl\{ e^{-\gamma_0 t}
    + \tau_0 \Bigl[1-e^{-\gamma_0 t}\Bigr] \biggl\}^2  \cr
& + \tau_0 {(1-\delta)^2\over 4\delta}
    \Bigl[1-e^{-\gamma_0 t}\Bigr] e^{-\gamma_0 t}                     \cr
& + {(1-\delta^2)^2\over 16\delta^2} \sin^4\Omega t
    e^{-2\gamma_0 t}                                                  \cr}
                                                        \eqno(3.3.12)
$$

\vskip 0.2cm

\noindent e) For weak damping, and for an initial minimum-uncertainty state,

$$
U_T(t)
= {\hbar^2\over 4}\biggl\{ e^{-\gamma_0 t}
+ \tau_0 \Bigl[1-e^{-\gamma_0 t}\Bigr] \biggl\}^2
                                                        \eqno(3.3.13)
$$

\noindent It is seen that the first term is of purely quantum nature
whereas the second term is of thermal nature, their contributions to
the uncertainty of the system arise from quantum and thermal
fluctuations respectively.

\vskip 0.5cm

\noindent{\bf 3.4 Weak Damping Limit }

\vskip 0.25cm

At arbitrary but finite temperatures,
the noise kernel (3.1.4) is not a Dirac delta
function, and in general there is no closed form for the uncertainty function.
However, if the damping is weak, we can find an approximate
expression for the uncertainty function.

Assuming that
$$
\alpha \equiv {\gamma_0\over 2\Omega } << 1
                                                        \eqno(3.4.1)
$$

\noindent then

$$
{{1\over 2}\gamma_0 \over
({1\over 2}\gamma_0)^2+(\Omega-\omega)^2}
\sim \delta(\Omega-\omega),
                                                        \eqno(3.4.2)
$$

\noindent which suggests that the major contributions to the integrals in
(A.4a-c) come from a small region near $\omega=\Omega$.
Using Eqs. (A.14) for (3.1.16-18), we obtain
$$
f_1(t) \sim
{\gamma_0\over 2\Omega_0} \coth\epsilon_r~
\sin^2\Omega t ~ e^{-\gamma_0 t}
                                                          \eqno(3.4.3)
$$
$$
g_1(t)
\sim {1\over 2}\coth\epsilon_r~
\biggl\{
   {{1+\delta^2}\over 2\delta}
   \Bigl[1-e^{-\gamma_0 t}\Bigr]
 - {{1-\delta^2}\over 2\delta} \alpha
   \sin 2\Omega t \biggr\} ~
   e^{-\gamma_0 t}
                                                         \eqno(3.4.4)
$$
and
$$
g_2(t)
\sim \Bigl[{1 \over 2}\coth\epsilon_r ~
(1-e^{-2\gamma_0 t} ) \Bigr]^2
                                                         \eqno(3.4.5)
$$

\noindent The uncertainty function (2.3.20) is then given by

$$
\eqalign{
{1\over\hbar^2}U_T(t)
& \sim {1 \over 4}\biggl\{ e^{-\gamma_0 t}
    +\coth\epsilon_r
     \Bigl[1-e^{-\gamma_0 t}\Bigr] \biggl\}^2                   \cr
& + \coth\epsilon
    \biggl\{ {(1-\delta)^2\over 4\delta}
    \Bigl[1-e^{-\gamma_0 t}\Bigr]
  -  {{1-\delta^2}\over 4\delta}
    \alpha~ \sin2\Omega t~  \biggr\}e^{-\gamma_0 t}          \cr
& +  \biggl\{ {{1-\delta^2}\over 4\delta} \sin 2\Omega t
  +  \alpha \Bigl[\coth\epsilon_r
    -{{1+\delta^2} \over 2\delta} \Bigr]
     \sin^2\Omega t \biggr\}^2 e^{-2\gamma_0 t}               \cr}
                                                        \eqno(3.4.6)
$$
\noindent This is the second main result of this paper.
One can deduce various limits from this expression.
For example, it is obvious that again at $t=0$,
when the initial uncorrelated conditions is assumed valid,
$ U_T(0)=\hbar^2/4$, which is the Heisenberg relation (1.1).
At very long time ($t>>\gamma_0^{-1}$), $ U_T(t) $ is insensitive to  $\delta$
and approaches $ U^{BE}_T $ as in (1.3) at finite temperature (as $\Omega
\simeq \Omega_0$).
That means the Brownian particle approaches an equilibrium
quantum statistical system. (For supraohmic bath this may not always be true).
We can also see that for a  $T=0$ bath ($\coth \epsilon =1$),
$U_T(t)$ has a leading term given
by $\hbar^2/4$ (the Heisenberg relation) followed, for squeezed states
$\delta \neq 1$, by terms of order $\alpha^0$ and
$\alpha$ depicting both decay and oscillatory behavior.
This is, of course, due to the action of quantum fluctuations alone.

For a minimum- uncertainty initial state ($\delta=1$),
we get for all finite temperatures
$$
U_T(t)
={\hbar^2\over 4}\biggl\{ e^{-\gamma_0 t}
+ \coth\epsilon
 \Bigl[1-e^{-\gamma_0 t}\Bigr] \biggl\}^2 + O (\alpha^2) ~~~(weak~~coupling)
%+ \hbar^2 \Bigl[ \alpha (\coth \epsilon -1) \sin^2\Omega t \Bigr]
% e^{-2 \gamma_0 t}
                                                        \eqno(3.4.7)
$$
This result was obtained in [5] earlier. Obviously, taking the high
temperature limit of (3.4.7) reduces to (3.3.13) above.

Notice that in (3.4.7) there is no linear order damping term. In this
calculation
we introduced a cutoff frequency $\Lambda$ in the spectral density $I(\omega)
\sim \omega \exp (-\omega^2 / \Lambda^2)$ which leads to a divergent term
dependent on $\Lambda$. It is removed by a frequency renormalization
procedure standard in field theory. (See Eq. (A.12c) and discussions in the
1992 paper of [15] and Appendix B of the 1983 paper of [13]).

At short times ($t<< \gamma_0^{-1}$),
$$
U_T(t)
\simeq {\hbar^2\over 4} \Bigl[ 1 + 2(\delta \coth {\hbar \Omega \over
2kT} - 1) \gamma_0 t + O (t^2) \Bigr]
                                                        \eqno(3.4.8)
$$
This simple  expression is revealing in several aspects:
The first term is the ubiquitous
quantum fluctuation, the second term is the thermal contribution, which depends
on the initial spread and increases with increasing dissipation and
temperature.
The time when thermal fluctuations overtake quantum fluctuations is
%when the
%second term in the square bracket becomes larger than unity which occurs at
(assuming the temperature is higher than the ground state energy):
$$
t_1 = {1 \over {2 \gamma_0 (\delta \coth { \hbar \Omega \over 2kT} - 1)}}
                                                        \eqno(3.4.9)
$$
\noindent Even though this result has not appeared in decoherence studies of
quantum to classical transitions [15],
we expect this to be equal to  the decoherence time scale $t_{dec}$
calculated for weak coupling for all temperatures.

\vfill
\eject

\centerline{\bf 4. Discussion}

%Now we use the results we obtained to discuss the questions we raised
%in the Introduction: i.e., the relation of quantum and
%thermal fluctuations and the relative role they play in quantum open systems.
%We discuss how this bears on two related issues: One
%on the relation of uncertainty to decoherence studies, and the other
%on the demarkation of quantum, classical and thermal regimes.
 Combining the generalized uncertainty relations obtained here
with the recent findings of the environment-induced decoherence
studies based on the same open-system framework,
where two characteristic times---the decoherence time $t_{dec}$ and the
relaxation time $t_{rel}$---are defined, one can address two basic issues
in quantum mechanics and in (nonequilibrium) quantum statistical mechanics
i.e.,
1) expound the relation between quantum, thermal and classical:
more specifically, depict the role of quantum and thermal fluctuations
in the quantum to classical transition; and
2) explicate the physical conditions for
 the realization of the two basic tenets of (equilibrium) quantum statistical
mechanics from quantum dynamics.

%For example,
%one can ask when the two postulates of equilibrium statistical mechanics
%are satisfied? When the thermal fluctuation overrides the quantum
%%fluctuations?
%and when the system assumes classical behavior? They can be seen to be
%related to each other and largely determined by the two time scales.

Quantum statistical mechanics of a macroscopic system is derived from
the quantum dynamics of its microscopic constituents under two basic
postulates [18]: i) random phase, and ii) equal {\it a priori} probability.
The first condition enables one to assign probability distributions to
a system occupying certain quantum states. It requires the suppression
of interference terms in the wave function or that the reduced density matrix
of the system be approximately diagonal. The second condition is the basis
for the microcanonical ensemble. When applied to the stipulations of
a canonical ensemble, it ensures that under general conditions
the system when put in contact with a large bath would eventually
equilibrate with it. We want to examine the processes by which
these two conditions are
attained from a more basic level, starting with the microdynamics of
a system of quantum particles. Specifically, we want to see if there is
a characteristic time  when the phase information is lost (Postulate i)
and another time when the system attains equilibrium with its surrounding
so that all accesible states (in the combined system and environment)
are equally probable (Postulate ii).

Now for the first issue which embodies the question we raised in the beginning
and which underlies the premises of the second issue:
%In loose terms, it is common practice to identify the high temperature
%regime of a system as the classical domain.
On the one hand, one often regards
the regime when thermal fluctuations begin to surpass quantum fluctuations
as the transition point from quantum to classical. On the other hand, from the
wave picture of quantum mechanics we know that a necessary condition for a
system to behave classically is that the interference terms in its wave
function have to diminish below a certain level,
%
%reduced density matrix to become approximately diagonal
%
so that probability can be assigned to classical events [2]
%
%in the decoherent history viewpoint, for the decoherent functional to assume
%an approximately diagonal form
%
or that classical decoherent histories can be well-defined [3]. This is
known as the decoherence process.
One should ask if there exists any relation between these two criteria of
classicality -- one based on the uncertainty relation and
the other on decoherence. In particular we want to see how the uncertainty
function $U_T(t)$ changes from the initial quantum fluctuation-dominated
condition to a later thermal fluctuation-dominated condition.
In Eq. (3.3.13) we show that under the stipulated  conditions  they
are indeed equivalent: The time the quantum system `decoheres' is also the time
when thermal fluctuation overtakes quantum fluctuations.

However, we issue a warning here that the regime after thermal fluctuations
dominate should not be called classical,
as is customary in many quantum to classical transition studies. In fact,
after the decoherence time only the first postulate of quantum statistical
mechanics (QSM) is satisfied, the system can be described by {\it
non-equilibrium}
QSM. Only after the relaxation time, when the second postulate is satisfied,
can the system be correctly described by {\it equilibrium} QSM.
The classical regime starts at a much later
time and it would have to be at a sufficiently high temperature.
It is well-known that quantum statistical effects can be important at very high
temperatures (e.g., Fermi temperature for metals). This is due to exchange
interactions of identical particles, a distinctly quantum effect. Only when
the statistical properties of fermions and bosons can be approximated by
distinguishable particles, usually at high temperatures when the Fermi-Dirac
or Bose-Einstein statistics approaches the Maxwell-Boltzmann statistics, can
the system be rightfully called {\it classical}. In this regard {\it quantum}
carries two meanings, one refers to the interference effect and the other
refers
to spin-statistics effect.

%The usage of the word classical in many decoherence studies
%is for practical considerations incorrect.

%One can also see from this calculation the time-development of
%the system from the initial quantum to the final classical regimes.
%One can also stipulate the conditions that the basic hypothesis i) and ii)
%are satisfied, or , more interestingly, when they fail.
%In our studies
%the low temperature weak coupling and supraohmic environment may
%invalidate these postulates, which implies that quantum statistic mechanics
%cannot be applied to the descriptions of such systems.
%\end

The analysis of nonohmic environments will be reported later [19].
There one expects to see some qualitatively new
behavior in the case of supraohmic spectral density function and
at low temperatures, when quantum fluctuations can play a more determinant
role which can make the decoherence time longer and the relaxation
of the system to equilibrium more difficult to achieve.

\vskip .5 cm

\noindent{ \bf Acknowledgement}

We thank Juan Pablo Paz, Alpan Raval
for discussions and Jonathan Halliwell for explaining his information-
theoretic results. This work is partially supported by the National Science
Foundation under grant PHY91-19726.

\vfill
\eject

\centerline{\bf Appendix}

\vskip 0.25cm

Let us define

$$
\eqalign{
[ss]
& = {1\over 2} \int\limits_0^t ds_1 \int\limits_0^t ds_2 ~
    \sin\Omega s_1 ~ \sin\Omega s_2 ~
    e^{-{1\over 2}\gamma_0 (s_1+s_2)} ~                    \cr
& \times
    \gamma_0 \int\limits_0^{+\infty} {d\omega\over\pi}
    \omega \coth \epsilon (\omega)
     \cos\omega(s_1-s_2)                                   \cr }
                                                            \eqno(A.1a)
$$

$$
\eqalign{
[sc]
& = {1\over 2} \int\limits_0^t ds_1 \int\limits_0^t ds_2 ~
    \sin\Omega s_1 ~ \cos\Omega s_2 ~
    e^{-{1\over 2}\gamma_0 (s_1+s_2)} ~                    \cr
& \times
    \gamma_0 \int\limits_0^{+\infty} {d\omega\over\pi}
    \omega\coth \epsilon (\omega)
     \cos \omega(s_1-s_2)                                   \cr }
                                                            \eqno(A.1b)
$$

$$
\eqalign{
[cc]
& = {1\over 2} \int\limits_0^t ds_1 \int\limits_0^t ds_2 ~
    \cos\Omega s_1 ~ \cos\Omega s_2 ~
    e^{-{1\over 2}\gamma_0 (s_1+s_2)} ~                    \cr
& \times
    \gamma_0 \int\limits_0^{+\infty} {d\omega\over\pi}
    \omega \coth \epsilon (\omega)
     \cos\omega(s_1-s_2)                                   \cr }
                                                            \eqno(A.1c)
$$

\noindent then

$$
\eqalignno{
a_{11}(t)
& = {1\over \sin^2\Omega t}~ e^{\gamma_0 t} ~ [ss]
                                                              &(A.2a)\cr
a_{12}(t) + a_{21}(t)
& = 2{1\over \sin^2\Omega t}~ e^{{1\over 2}\gamma_0 t}
   \Bigl\{ \sin\Omega t ~ [sc] -\cos\Omega t ~ [ss] \Bigr\}
                                                              &(A.2b)\cr
a_{22}(t)
& = {1\over\sin^2\Omega t} \Bigl\{
    \cos^2\Omega t ~ [ss]
  - \sin2\Omega t  ~ [sc]
  + \sin^2\Omega t ~ [cc] \Bigr\}
                                                              &(A.2c)\cr }
$$

\noindent By switching the orders of the integrations in (A.1a), (A.1b)
and (A.1c), and by first performing the integrations of $s_1$ and $s_2$,
we obtain

$$
\eqalign{
[ss]
& = {\Omega^2\over 2\Omega_0^2}
    \biggl\{I_{\gamma c}(0)+\alpha
     I_{\omega c}(0)\biggr\}                                            \cr
& - {\Omega^2\over 2\Omega_0^2}\biggl\{
     \Bigl[\cos\Omega t+
    \alpha\sin\Omega t\Bigr]
     \Bigl[I_{\gamma c}(t)
   +\alpha I_{\omega c}(t)\Bigr]
   +{\Omega_0^2\over\Omega^2}I_{\gamma s}(t) \biggl\}
    e^{-{1\over 2}\gamma_0 t}                                          \cr
& + {\Omega^2\over 2\Omega_0^2}\biggl\{
    \Bigl[{\Omega_0^2\over\Omega^2}
   -\alpha^2\cos 2\Omega t
   +\alpha\sin 2\Omega t \Bigr]I_{\gamma c}(0)    \cr
&  +\alpha\Bigl[\cos 2\Omega t
   +\alpha\sin 2\Omega t\Bigr]
   I_{\omega c}(0) \biggr\} e^{-\gamma_0 t}                             \cr }
                                                            \eqno(A.3a)
$$

$$
\eqalign{
[sc]
& = {\Omega^2\over 2\Omega_0^2} \alpha
    \biggl\{I_{\gamma c}(0)+\alpha
     I_{\omega c}(0)\biggr\}                                           \cr
& - {\Omega^2\over 2\Omega_0^2}\alpha
    \biggl\{ \Bigl[\cos\Omega t
  + \alpha\sin\Omega t\Bigr] I_{\gamma c}(t)
  + \Bigl[\alpha\cos\Omega t
   -\sin\Omega t\Bigr] I_{\omega c}(t)
    \biggr\}e^{-{1\over 2}\gamma_0 t}                                 \cr
& + {\Omega^2\over 2\Omega_0^2} \alpha
    \biggl\{\Bigl[\cos 2\Omega t
   +\alpha\sin 2\Omega t \Bigr]
    I_{\gamma c}(0)                                                  \cr
&  +\alpha
    \Bigl[\alpha\cos 2\Omega t
    -\sin 2\Omega t\Bigr]
     I_{\omega c}(0) \biggr\} e^{-\gamma_0 t}                          \cr }
                                                            \eqno(A.3b)
$$

$$
\eqalign{
[cc]
 & = {\Omega^2\over 2\Omega_0^2} \biggl\{
     \Bigl[1+{\gamma_0^2\over 2\Omega^2}\Bigl] I_{\gamma c}(0)
     -\alpha I_{\omega c}(0)\Bigr\}                   \cr
 & + {\Omega^2\over \Omega_0^2}\biggl\{
      \Bigl[\alpha\sin\Omega t
     -\Bigl(1+{\gamma_0^2\over 2\Omega^2}\Bigl) I_{\gamma c}(t)
     +\alpha\Bigl( \cos\Omega t
     +\alpha\sin\Omega t\Bigr) I_{\omega c}(t)     \cr
 &   -{\Omega_0^2\over \Omega^2}
      \sin\Omega t  I_{\gamma s}(t)
      \biggl\} e^{-{1\over 2}\gamma_0 t}                               \cr
&  + {\Omega^2\over 2\Omega_0^2} \biggl\{
      \Bigl[{\Omega_0^2\over\Omega^2}
    +\alpha^2\cos 2\Omega t
    -\alpha\sin 2\Omega t\Bigr]
     I_{\gamma c}(0)                                                   \cr
&   +\alpha\Bigl[-\cos 2\Omega t
     -\alpha\sin 2\Omega t\Bigr]
     I_{\omega c}(0) \biggr\} e^{-\gamma_0 t}                          \cr }
                                                            \eqno(A.3c)
$$

\noindent where

$$
I_{\gamma c}(t)
={1\over 2}\int\limits_0^{+\infty}{d\omega\over\pi}
\omega\coth\epsilon(\omega)
\Bigl[\gamma_{-}+\gamma_{+}\Bigr]~\cos\omega t
                                                             \eqno(A.4a)
$$

$$
I_{\gamma s}(t)
={1\over 2}\int\limits_0^{+\infty}{d\omega\over\pi}
\omega\coth\epsilon(\omega)
\Bigl[\gamma_{-}-\gamma_{+}\Bigr]~\sin\omega t
                                                             \eqno(A.4b)
$$

$$
I_{\omega c}(t)
={1\over 2}\int\limits_0^{+\infty}{d\omega\over\pi}
\omega\coth\epsilon(\omega)
\Bigl[\omega_{-}+\omega_{+}\Bigr]~\cos\omega t
                                                             \eqno(A.4c)
$$

\noindent and

$$
\gamma_{\pm}
={ {1\over 2}\gamma_0 \over
({1\over 2}\gamma_0)^2 +(\Omega \pm \omega)^2 }
                                                             \eqno(A.5a)
$$

$$
\omega_{\pm}
={  \Omega \pm \omega \over
({1\over 2}\gamma_0)^2 + (\Omega \pm \omega)^2 }
                                                             \eqno(A.5b)
$$

\vskip 0.2cm

\noindent (a) In the {\it high temperature} limit

$$
I_{\gamma c}(t)
\simeq {kT\over \hbar}\int\limits_0^{+\infty}{d\omega\over\pi}
       \Bigl[\gamma_{-}+\gamma_{+}\Bigr]~\cos\omega t
     = {kT\over \hbar} e^{-{1\over 2}\gamma_0 t} \cos\Omega t
                                                             \eqno(A.6a)
$$

$$
I_{\gamma s}(t)
\simeq {kT\over \hbar}\int\limits_0^{+\infty}{d\omega\over\pi}
       \Bigl[\gamma_{-}-\gamma_{+}\Bigr]~\sin\omega t
     = {kT\over \hbar} e^{-{1\over 2}\gamma_0 t} \sin\Omega t
                                                             \eqno(A.6b)
$$

$$
I_{\omega c}(t)
\simeq {kT\over \hbar}\int\limits_0^{+\infty}{d\omega\over\pi}
       \Bigl[\omega_{-}+\omega_{+}\Bigr]~\cos\omega t
     = {kT\over \hbar} e^{-{1\over 2}\gamma_0 t} \sin\Omega t
                                                             \eqno(A.6c)
$$

\noindent In particular,

$$
I_{\gamma c}(0) \simeq {kT\over \hbar}
                                                             \eqno(A.7a)
$$

$$
I_{\omega c}(0) \simeq 0
                                                             \eqno(A.7b)
$$

\noindent then

$$
\eqalignno{
[ss]
& \simeq {kT\over \hbar} {\Omega^2\over 2\Omega_0^2}
\biggl\{ 1 - \Bigl[ {\Omega_0^2\over\Omega^2}
-\alpha^2\cos 2\Omega t
+\alpha\sin 2\Omega t \Bigr]
e^{-\gamma_0t} \biggr\}
                                                            &(A.8a) \cr
[sc]
& \simeq {kT\over \hbar} {\Omega^2\over 2\Omega_0^2}
\alpha
\biggl\{ 1 - \Bigl[\cos 2\Omega t
+\alpha\sin 2\Omega t \Bigr]
e^{-\gamma_0t} \biggr\}
                                                            &(A.8b) \cr
[cc]
& \simeq {kT\over \hbar} {\Omega^2\over 2\Omega_0^2}
\biggl\{ \Bigl[1+{\gamma_0^2\over 2\Omega^2}\Bigr]
-\Bigl[ {\Omega_0^2\over\Omega^2}
+\alpha^2\cos 2\Omega t
-\alpha\sin 2\Omega t \Bigr]
e^{-\gamma_0 t} \biggr\}
                                                            &(A.8c) \cr }
$$

\noindent Substituting Eq. (A.8a-c) into Eq. (3.1.17), (3.1.18) and (3.3.19),
we obtain Eq. (3.3.1), (3.3.2) and (3.3.3).

\vskip 0.2cm

\noindent (b) For {\it weak damping},

If the damping constant is very small, i.e.,

$$
\alpha << 1
                                                           \eqno(A.9)
$$

\noindent then

$$
{{1\over 2}\gamma_0 \over
({1\over 2}\gamma_0)^2+(\Omega-\omega)^2}
\sim \delta(\Omega-\omega),
                                                           \eqno(A.10)
$$

\noindent which suggests that the major contributions to integrals (A.4a),
(A.4b) and (A.4c) come from a small region near $\omega=\Omega$.  Thus we
can make the following expansion:

$$
{\omega\over\tanh{\hbar\omega\over 2k_BT}} \sim
{\Omega\over\tanh\epsilon_r}
\biggl\{1-\Bigl[1-{\epsilon_r\over
\sinh\epsilon_r}\Bigl]
{\Omega-\omega \over \Omega } \biggr\}
                                                        \eqno(A.11)
$$

\noindent to get

$$
I_{\gamma c}(t)
\sim {{1\over 2}\Omega\over\tanh\epsilon_r}~
e^{-{1\over 2}\gamma_0 t}~\cos\Omega t
\biggl\{1-\alpha
\Bigl[1-{\epsilon_r\over
\sinh\epsilon_r}\Bigl]
{\sin\Omega t \over \cos\Omega t } \biggr\}
                                                       \eqno(A.12a)
$$

$$
\eqalign{
I_{\gamma s}(t)
& \sim {{1\over 2}\Omega\over\tanh\epsilon_r}~
e^{-{1\over 2}\gamma_0 t}~\sin\Omega t
\biggl\{1+\alpha
\Bigl[1-{\epsilon_r\over
\sinh\epsilon_r}\Bigl]
{\cos\Omega t \over \sin\Omega t } \biggr\} \cr
& -{\sqrt{\pi}\over\tanh\epsilon_r}
\Bigl[1-{\epsilon_r\over\sinh\epsilon_r}\Bigr]
\sqrt{\Lambda}e^{-{1\over 4}\Lambda^2t^2} \cr}
                                                       \eqno(A.12b)
$$
The second term drops out when a high frequency cutoff $\Lambda$ is assumed.
$$
I_{\omega c}(t)
\sim {{1\over 2}\Omega\over\tanh\epsilon_r}~
e^{-{1\over 2}\gamma_0 t}~\sin\Omega t
\biggl\{1-\alpha
\Bigl[1-{\epsilon_r\over
\sinh\epsilon_r}\Bigl]
{\sin\Omega t \over \cos\Omega t } \biggr\}
                                                       \eqno(A.12c)
$$

\noindent and

$$
I_{\gamma c}(0)
\sim {{1\over 2}\Omega\over\tanh\epsilon_r}~
                                                       \eqno(A.13a)
$$

$$
I_{\omega s}(0)
\sim {{1\over 2}\Omega\over\tanh\epsilon_r}~
\alpha
\Bigl[1-{\epsilon_r\over
\sinh\epsilon_r}\Bigl]
                                                       \eqno(A.13b)
$$

\noindent Substituting (A.12a-c) and (A.13a-b) into (A.3a-c), we obtain

$$
[ss]
\simeq {\Omega\over 4}\coth \epsilon_r
\biggl\{1-\Bigl[1+\alpha\sin 2\Omega t \Bigr]
e^{-\gamma_0 t} \biggr\}
                                                       \eqno(A.14a)
$$

$$
[sc]
\simeq {\Omega\over 4}\coth \epsilon_r
\alpha
\biggl\{ 1 - \Bigl[\cos 2\Omega t
+\alpha\sin 2\Omega t \Bigr]
e^{-\gamma_0t} \biggr\}
                                                       \eqno(A.14b)
$$

$$
[cc]
\simeq {\Omega\over 4}\coth \epsilon_r
\biggl\{1-\Bigl[1-\alpha\sin 2\Omega t \Bigr]
e^{-\gamma_0 t} \biggr\}
                                                       \eqno(A.14c)
$$

\vfill
\eject

\noindent {\bf References}

\noindent [1] See., e.g.,  K. Lindenberg and B. J. West, {\it The
Nonequilibrium
Statistical Mechanics of Open and Closed Systems} (VCH, New York, 1990)
%E. B. Davies, {\it Quantum Theory of Open Systems}( Academic Press, New York,
%%1976)

\noindent [2] W. H. Zurek, Phys. Rev. D24, 1516 (1981); D26, 1862 (1982);
    in {\it Frontiers of Nonequilibrium Statistical Physics},
    ed. G. T. Moore and M. O. Scully (Plenum, N. Y., 1986);
    E. Joos and H. D. Zeh, Z. Phys. B59, 223 (1985);
    A. O. Caldeira and A. J. Leggett, Phys. Rev. A31, 1059 (1985)
    W. G. Unruh and W. H. Zurek, Phys. Rev. D40, 1071 (1989);
    W. H. Zurek, Prog. Theor. Phys. 89, 281 (1993)

\noindent [3] M. Gell-Mann and J. B. Hartle, in {\it Complexity, Entropy
             and the Physics of Information}, ed. W. Zurek,
             Vol. IX (Addison-Wesley, Reading, 1990);
             Phys. Rev. D47, 3345 (1993);
             R. Griffiths, J. Stat. Phys. 36, 219 (1984);
             R. Omnes, Rev. Mod. Phys. 64, 339 (1992);
             H. F. Dowker and J. J. Halliwell, Phys. Rev. D46, 1580 (1992);
             T. Brun, Phys. Rev. D47, 3383 (1993);
             J. B. Hartle, in {\it Directions in General Relativity, Vol 1:
             Misner Festschrift}, eds. B. L. Hu, M. P. Ryan and C. V.
             Vishveswara (Cambridge University Press, Cambridge 1993);
             E. Calzetta and B. L. Hu, in {\it Directions in General
             Relativity, Vol 2: Brill Festschrift},
             eds. B. L. Hu and T. A. Jacobson
             (Cambridge University Press, Cambridge 1993)

\noindent [4] B. L. Hu and Yuhong Zhang, in {\it Quantum Dynamics of Chaotic
             Systems} eds. J. M. Yuan, D. H. Feng and G. M. Zaslavsky
              (Gordon and Breach, Langhorne, 1993).

\noindent [5] B. L. Hu and Yuhong Zhang, "Uncertainty Relation at Finite
Temperature"  University of Maryland preprint umdpp 93-161 (1992)

\noindent [6] I. Bialynicki-Birula and J. Mycielski, Comm. Math. Phys. 44,
    129 (1975); W. Beckner, Ann. Math. 102, 159 (1975).

\noindent [7]    D. Deutsch, Phys. Rev. Lett. 50, 631 (1983);
    M. H. Partovi, Phys. Rev. Lett. 50, 1883 (1983);
    See also S. Abe and N. Suzuki, Phys. Rev. A41, 4608 (1990)

\noindent [8] A. Anderson and J. J. Halliwell, Phys. Rev. D48, 2753 (1993)

\noindent [9] J. J. Halliwell,  Phys. Rev. D48, 2739 (1993)

\noindent [10] A. Mann, M. Revzen, H.Umezawa and Y. Yamanaka, Phys. Lett.
              A140, 475 (1989)

\noindent [11] R. Rubin, J. Math. Phys. 1, 309 (1960);
              J. Schwinger, J. Math. Phys. 2, 407 (1961);
              G. W. Ford, M. Kac, P. Mazur, J. Math. Phys. 6, 504 (1963);
              %H. Dekker, Phys. Rev. A16, 2116 (1977);
              H. Dekker, Phys. Rep. 80, 1 (1981);
              V. Hakim and V. Ambegoakar, Phys. Rev. A36, 3509 (1985);
              F. Haake and R. Reibold, Phys. Rev. A32, 2462 (1985).

\noindent [12] R. P. Feynman and F. L. Vernon, Ann. Phys. 24, 118 (1963)

\noindent [13] A. O. Caldeira and A. J. Leggett, Physica 121A, 587 (1983);
    A. J. Leggett et al, Rev. Mod. Phys. 59, 1 (1987)

\noindent [14] H. Grabert, P. Schramm, and G. -L. Ingold,
                     Phys. Rep. 168, 115 (1988)

\noindent [15] B. L. Hu, J. P. Paz and Y. Zhang, Phys. Rev. D45, 2843 (1992);
               D47, 1576 (1993)

\noindent [16]  J. P. Paz, S. Habib and W. H. Zurek, Phys. Rev. D47, 488
(1993);
    W. H. Zurek, S. Habib and J. P. Paz, Phys. Rev. Lett. 47, 1187(1993);
    J. P. Paz and W. H. Zurek, Phys. Rev. 48, 2728 (1993).

\noindent [17] See, e.g., B. L. Schumacher, Phys. Rep. 135, 317 (1986).
    For recent work, see, e.g., Y. S. Kim and W. Zachary, eds
    {\it Squeezed State and  Uncertainty Relation} (NASA Publication, 1992)

\noindent [18] See, e.g., K. Huang, {\it Statistical Mechanics}, 2 ed. (Wiley,
    New York, 1987)

\noindent [19] B. L. Hu, A. Raval and Y. Zhang, in preparation
\end